\documentclass[natbib, envcountsect]{svjour3}

\usepackage{graphicx}
\journalname{Journal on Data Semantics}

\usepackage[T1]{fontenc}
\usepackage[utf8]{inputenc}

\let\vec\relax 
\DeclareMathAccent{\vec}{\mathord}{letters}{"7E} 

\usepackage{amsmath}
\usepackage{amssymb}
\usepackage{mathtools}
\usepackage[mathscr]{euscript}
\usepackage{url}
\usepackage{siunitx}
\usepackage{booktabs}
\usepackage{mdframed}
\usepackage{diagbox}
\usepackage{tabularx}
\usepackage{ragged2e}
\usepackage{makecell}
\usepackage{slashbox}
\usepackage[linesnumbered,ruled,vlined]{algorithm2e}

\usepackage[noend]{algpseudocode}
\usepackage[section]{placeins} 

\usepackage{tikz}

\usepackage{pgfplots}
\pgfplotsset{width=7cm,compat=newest}
\usetikzlibrary{patterns}

\newcommand{\tabitemplus}{~\llap{\textbf{+}}~}
\newcommand{\tabiteminus}{~\llap{\textbf{--}}~}
\SetCommentSty{mycommfont}

\SetKwInput{KwInput}{Input}                
\SetKwInput{KwOutput}{Output}              
\algnewcommand\AND{\textbf{and }}
\algnewcommand\OR{\textbf{or }}
\algnewcommand\NOT{\textnormal{\textbf{not }}}
\algnewcommand\TRUE{\textnormal{\textbf{true}}}
\algnewcommand\FALSE{\textnormal{\textbf{false }}}
\algnewcommand\CONCAT{\textnormal{\textbf{concat }}}

\renewcommand{\vec}[1]{\mathbf{#1}} 

\newcommand*{\orcid}[1]{\href{https://orcid.org/\detokenize{#1}}{\includegraphics[scale=1]{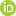}}}

\usepackage[bookmarks=false]{hyperref}
\usepackage[nameinlink,noabbrev]{cleveref}

\begin{document}


\title{Geo-L:  Linking Geospatial Data Made Easy \thanks{This work has been supported by the DataBio project,  funded by the European Union's Horizon 2020 research and innovation programme under grant agreement No. 732064.}
}


\author{Christian Zinke-Wehlmann 
\and
        Amit Kirschenbaum 
}

\authorrunning{C. Zinke-Wehlmann and A. Kirschenbaum}


\institute{Institut f\"ur Angewandte Informatik an der Universit\"at Leipzig (InfAI)\\
Goerdelerring 9\\
04109 Leipzig, Germany\\
\email{zinke@infai.org} \\
\email{amit@informatik.uni-leipzig.de}}

\date{}

\maketitle
\begin{abstract}
Geospatial Linked Data is an emerging domain, with growing interest in research and industry. 
There is an increasing number of publicly available geospatial Linked Data resources and they need to be interlinked and easily integrated with private and industrial Linked Data on the Web. 
The present paper introduces Geo-L, a system for discovery of RDF spatial links based on topological relations. Experiments show that the proposed system improves state-of-the-art spatial linking processes in terms of mapping-time and -accuracy, as well as concerning resources retrieval efficiency and robustness.

\end{abstract}

\section{Introduction}
\label{sec:introduction}
Web of Data, or Semantic Web, is a continuously growing global data space.\footnote{see: https://www.w3.org/2013/data/} Semantic Web standards, such as RDF \citep{klyne2004rdf,rdf-wg2014rdf}, OWL \citep{bechhofer2004owl,owl-wg2012owl}, and SPARQL \citep{prudhommeaux2008sparql} were developed to express and exchange semantic information on the Web, which tackle the challenge of interoperability \citep{hitzler2009foundations}. In the geospatial context, most prominent is the GeoSPARQL initiative, which offers a necessary vocabulary to develop geo-related data on the Semantic Web \citep{battle2011geosparql}. In recent years,  geospatial linked data gained increasing attention \citep{NIKOLAOU201535}, also due to advances in the Earth Observation domain \citep{7994582}. Thus, numerous resources of linked geospatial data have been developed, e.g., LinkedGeoData \citep{auer2009linkedgeodata}, Smart Point Of Interest \citep{vcerba2016sdi4apps}, Spanish Cases \citep{de2010geographical}, and Ireland's national geospatial data \citep{debruyne2016serving}; the domain is constantly growing within the Linked Data Cloud. Notably, the domain of geospatial data contains complex datasets, as NUTS \citep{eurostat2015regions}, which describe territories using polygons that may be more than 1700 points long.

According to the Linked Data principles, published data should be interlinked with other datasets on the Web \citep{bizer2011linked}. In general, linking (and fusing) of geospatial linked data sources enable large-scale inferences and data integration \citep{wiemann2016spatial-data-fusion}. Nevertheless, explicit links are often not part of the dataset and should be discovered automatically, even in a distributed cloud environment and huge datasets. These linking activities are one pillar to foster the development of innovative software solutions. In particular, the linking of geospatial data is a challenging task, since the links express relations which depend on complex geometric computations. 


The present work introduces Geo-L, a system for discovery of spatial links in RDF datasets according to topological relations. 
Geo-L was developed considering the following requirements, which we identified by comparing existing approaches, services, and tools for this task:

\begin{enumerate}
\item Scalability and efficiency: As mentioned before, the Linked Data cloud is continually growing employing new sources and data sets. The service should be able to handle big data sets. The idea is to provide a service for different Linked Data environments (open or closed). Therefore, the time efforts have to be reduced on a significant minimum. The vision is to discover even extensive data sets in near real-time.
\item Robustness: The service must retain functionality under unforeseen conditions, as missing or corrupted data. This is especially true for crowd-sourced or automatically generated data sets, which are likely to include errors as the size of data grows.
\item Interoperability and flexibility: The service has to be handled as easy and transparent as possible.  The (SPARQL affine) user should be able to easily formulate queries to retrieve source and target datasets,  as well as the linking condition. This includes the ability to handle data whose representation is not compatible for computing of, e.g., topological relations. Further, the service has to handle on-the-fly requests by a RESTful input processing. It has to operate easily as a standalone system or as a module integrated into other applications.  
\end{enumerate}

\section{Background}
Linked (Open) Data refer to an area which focuses on the publishing of RDF (Resource Description Framework) on the Web of Data. However, the Linked Data approach is strongly linked to the Linked Data Principles by Tim Berners-Lee \citep{bizer2011linked}. The basic idea of link discovery is to find data items  within the target dataset which are logically connected to the source dataset. More formaly this means: Given $\mathscr{S}$ and $\mathscr{T}$, sets of RDF resources, called source and target resources respectively,  and a relation $R$, the aim of link discovery methods is to find a mapping $M$ such that $M = \{(s, t) \in \mathscr{S} \times \mathscr{T} : R(s, t)\}$. Naive computation of M requires quadratic time complexity to test for every $s \in \mathscr{S}$ and  $t \in \mathscr{T}$ whether $R$ holds, which is unfeasible for large data sets. 

In geospatial context, $\mathscr{S}$ and $\mathscr{T}$ are sets of spatial objects, which contain geometries in a two dimensional space as features; the links may be based on proximity or on topological relations. In the latter case, relations are expressed by the Dimensionally Extended nine-Intersection Model (DE+9IM) \citep{clementini1993small,clementini1994modelling}, which was accepted as an ISO standard \citep{ISO19107}.
DE+9IM classifies binary spatial relationships between two geometries, $a$ and $b$, which may be points, lines, or polygons, based on intersection of interiors (I), boundaries (B) exteriors (E) of $a$ with those of $b$ .  

A combination of these six geometric features define topological relations, which are described in a $3 \times 3$ matrix as follows:
\[
\small{
\begin{array}{l}
\text{DE+9IM}(a,b) = \\
\begin{bmatrix}
dim(I(a)\cap I(b)) & dim(I(a)\cap B(b)) & dim(I(a)\cap E(b)) \\
dim(B(a)\cap I(b)) & dim(B(a)\cap B(b)) & dim(B(a)\cap E(b)) \\
dim(E(a)\cap I(b)) & dim(E(a)\cap B(b)) & dim(E(a)\cap E(b)) \\
\end{bmatrix}
\end{array}
}
\]
The intersection $S$ of some feature of $a$ with a feature of $b$, may be either empty or in itself a geometric object, namely: a point, a line, or a polygon. $dim(S)$ returns the dimension of the geometry $S$; if $S$ consists of multiple geometries then $dim(S)$ is the maximal dimension of  intersection if it is of multiple parts.  

\begin{equation*}
dim(S) = 
\left\{
\begin{aligned}
-1             & \quad\text{if } S=\varnothing\\
 \hphantom{-}0 & \quad\text{if $S$  contains at least one point,}\\                       & \quad\text{but no lines or polygons}\\
 \hphantom{-}1 & \quad\text{if $S$  contains at least one line,}\\
               & \quad\text{but no polygons}\\
 \hphantom{-}2 & \quad\text{ if $S$  contains at least one polygon}
\end{aligned}
\right.
\end{equation*}

In addition to the dimensions values the matrix may contain the values \texttt{T} $(dim(S)\geq 0)$, \texttt{F} $(dim(S) = -1)$, and \texttt{*} (``don't-care'' value, which means that the value in this matrix cell has no influence on the outcome of a function applied to this matrix).  
The model defines topological predicates to describe the spatial relations between the two geometries in a compact and human-interpretable manner, which are defined by pattern matrices:  
\emph{equals}, \emph{disjoint}, \emph{intersects}, \emph{touches}, \emph{crosses}, \emph{overlaps}, \emph{within}, and \emph{contains}.
For example, the pattern matrix for the relation \emph{within} is defined by the following pattern matrix \footnote{see also  \citet{strobl2008dimensionally}}
\begin{equation*}
    a.within(b) =
    \begin{bmatrix}
    \texttt{T} & \texttt{*} & \texttt{F} \\
    \texttt{*} & \texttt{*} & \texttt{F} \\
    \texttt{*} & \texttt{*} & \texttt{*} \\
    \end{bmatrix}
\end{equation*}
formally described as $(I(a)\cap I(b)\neq \varnothing) \wedge \neg(I(a)\cap E(b) \neq \varnothing) \wedge \neg(B(a)\cap E(b)\neq \varnothing)$.

\begin{figure}[ht]
    \centering
    \includegraphics[scale=0.25]{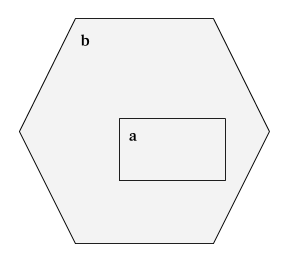}
    \caption{a within b}
    \label{fig:a_within_b}
\end{figure}

To illustrate how this matrix, and hence, the formula define the \emph{within} relation consider \Cref{fig:a_within_b}, which shows two geometries $a$ and $b$, such that $a$ is within $b$. We use \Cref{tab:example_within_DE+9IM_features} to graphically depict the respective  features $f_1(a),f_2(b)$, such that $f_1,f_2\in\{I,B,E\}$,  used in each component of the \emph{within} formula, for those two geometries, as well as the dimension of their intersection. 
As can be observed the conditions of the topological relation  \emph{within} are satisfied. 

\begin{table}
    \centering
    \begin{tabular}{llc}
    \toprule
    \multicolumn{2}{c}{$f_1(a),f_2(b)$} & $dim(f_1(a) \cap f_2(b))$\\  
    \cmidrule(l){1-2} \cmidrule(l){3-3} 
    $I(a), I(b)$ & \raisebox{-0.80\totalheight}{\includegraphics[scale=0.25]{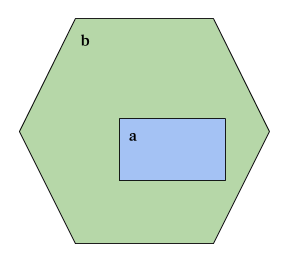}} & 2\\
    $I(a), E(b)$ &  \raisebox{-0.80\totalheight}{\includegraphics[scale=0.25]{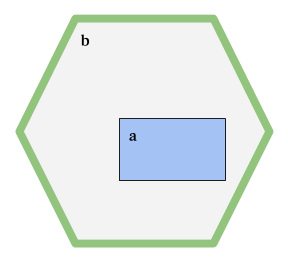}} & -1 \\
    $B(a), E(b)$ &  \raisebox{-0.80\totalheight}{\includegraphics[scale=0.25]{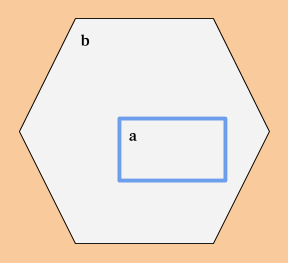}} & -1\\         
    \end{tabular}
    \caption{Geometry features of components of \emph{within} formula and dimension of their intersections }
    \label{tab:example_within_DE+9IM_features}
\end{table}

\section{Related Work}
Link discovery of topological relations among RDF data sets has received growing interest in recent years, and various methods for this problem have been proposed.  These methods usually define the topological relations between two geometries based on their relations computed between their minimum bounding boxes. A minimum bounding box (MBB) is the rectangle of minimum area that encloses all coordinates of geometry and is a commonly used as an approximation to the geometry to reduce computational costs that involve this geometry \citep{freeman1975determining}.

\citet{smeros2016discovering} use the \emph{MultiBlocking}  technique \citep{isele2011efficient} to discover topological relations. This technique divides the earth surface into curved rectangles, and assigns each geometry to all blocks in which it intersects, based on the geometry's MBB. 
Relations discovered within each block are then aggregated to construct the links. This method is embedded in the Silk framework \citep{volz2009silk}.

\textsc{Radon} \citep{sherif2017radon} divides the space into hyper-cubes and uses optimized sparse space tiling to index geometries. This is done by mapping each geometry to the set of hyper-cubes over which it's minimum bounding box (MBB) spans. The method first indexes geometries $s\in \mathscr{S}$  and then only index geometries $t \in \mathscr{T}$ that may potentially reside in hyper-cubes already contained in the index. To minimize the size of the index, the method implements a swapping strategy, that is, prior to the indexing phase it calculates an estimated total hypervolume ($eth$) for each of the datasets $\mathscr{S}$ and $\mathscr{T}$. If $eth(\mathscr{T})<eth(\mathscr{S})$ then it swaps the two datasets and computes the reverse relation of the requested relation $R$. The link generation itself is done using a method that reduces computations on a subset of DE+9IM relations.   \textsc{Radon} is implemented as part of the LIMES framework \citep{ngomo2011limes}\footnote{\url{https://github.com/dice-group/LIMES/}}.

\citet{faria2017results} adapt the AgreementMakerLight (AML) \citep{faria2014agreementmakerlight}, a framework for automated ontology matching, to tackle the task of  topological relations. This is done by utilizing ESRI Geometry API\footnote{\url{https://github.com/Esri/geometry-api-java/}}, which uses quadtree as means to index geometries and detect topological relationship among them. 

These methods, as well as OntoIdea \citep{khiat2017match}, were evaluated on several sets of geometries: \citet{achichi2017results} apply them to discover topological relations between LineStrings, constructed of  trajectories from the TomTom\footnote{\url{https://www.tomtom.com}} dataset.
\citet{saveta2018spgen}  apply these methods to find relations between LineStrings to LineStrings and between LineStrings to Polygons, from TomTom dataset and Spaten dataset \citep{doudali2017spaten}  respectively. 
All datasets included at most 2000 instances.
Both evaluations report that the methods mentioned above discover links correctly, that is, the $F$-score of most of them is $1.0$ (apart from OntoIdea which $F$-score lies between $0.91$ and $0.99$, and did not take part in the  tasks for link discovery between linestrings and polygons).

Strabon \citep{kyzirakos2012strabon} is an open-source geospatial RDF store. It is based on the RDF4J (previously Sesame) RDF store and adds geospatial capabilities to it by implementing the OGC-standard GeoSPARQL, where as part of the implementation the stored geometries in Strabon are indexed with an R-Tree-over-GiST.  Implementing GeoSPARQL means that Strabon includes topological functions; thus, queries that use these functions can be viewed as a means to discover topological relations.
\citet{sherif2017radon} compares the performance of Silk, Strabon, and \textsc{Radon} where they are applied to discover links between different subsets of NUTS and \emph{CORINE Land Cover}\footnote{\url{https://www.eea.europa.eu/publications/COR0-landcover}} datasets, which map land and land-usage respectively. The biggest dataset used in their experiments is of size $2,209,538$.           

The evaluations compare the running times of these methods with different dataset sizes. It has already been acknowledged that a significant portion big data is geospatial data \citep{lee2015geospatial,li2016geospatial}, thus our interest lies in the performance of these systems on large datasets. Table~\ref{tab:geospatial-linking-methods} summarizes  how well the methods described above perform, regarding the criteria for  useful geospatial link discovery systems, discussed in \Cref{sec:introduction}, as reported in the literature \citep{sherif2017radon,achichi2017results,saveta2018spgen}. 

\begin{center}
\begin{table*}
\begin{minipage}{\textwidth}
{\small
\hfill{}
\begin{tabularx}{\textwidth}{lXXX}
\toprule
\textbf{System} & \textbf{Scalability \& Efficiency} & \textbf{Robustness}  & \textbf{Interoperability \& Flexibility} \\
\midrule
Silk     &  \tabiteminus long running time on large data-sets & \tabiteminus instances limited to   size of 64K  & \tabitemplus standalone framework \\
&  & \tabiteminus not evaluated for relations \emph{cover} and \emph{covered by} &  \tabitemplus has REST and programmable APIs \\
& & & \tabiteminus linkage definition language is restricting \\
& & & \tabiteminus does not support transformation of geospatial data\\
\midrule
AML  & \tabitemplus achieves best run time for \emph{touches} and \emph{intersects} for LineStrings & \tabiteminus reaches time limit for \emph{disjoint} (75 min.) & \tabitemplus uses ESRI, an external module for handling geometries \\
     & \tabiteminus long running time on large data-sets for LineString/ Polygon tasks for \emph{contains} \emph{within} \emph{covers}  & \tabiteminus no information is given about error handling &  \tabiteminus strict linkage definition \\
\midrule
OntoIdea  &
\tabiteminus long running time on large datasets & \tabiteminus not evaluated for \emph{disjoint} & \tabiteminus no specification given \\
& \tabiteminus not evaluated for large data-sets &  \tabiteminus no information about error handling & \\
\midrule
Strabon & 
\tabitemplus run time for \emph{intersects} on smaller data-sets is better than that of LIMES & \tabiteminus did not finish any experiment on a large dataset within the time limit (2 hours) & \tabitemplus implements GeoSPAQRL, thus is able to transform geospatial object in retrieval time \\ 
 & & \tabiteminus doesn't provide feedback about progress of its task & \\
 & & \tabiteminus no transparent error handling & \\
\midrule
LIMES & \tabitemplus addresses all tasks regarding topological link discovery & \tabiteminus data or server error interrupt whole process & \tabitemplus can be applied as part of a framework or as a part of an application via its API \\
 & \tabitemplus achieves the best run-time performance for most of the topological relations (except intersect, and touches) &  & \tabiteminus  strict linkage definition (XML), no direct SPARQL support \\
\midrule
Geo-L & \tabitemplus addresses all tasks regarding topological link discovery & \tabitemplus storing chunks of datasets regularly minimizes data loss if connection is interrupted due to e.g., server error & \tabitemplus  can be applied as an independent application or through its API (as well as via REST API) \\
      & \tabitemplus achieves the best run-time performance for all topological relations & \tabitemplus provides feedback about task progress & \tabitemplus supports dataset definition via SPARQL query  \\ 
\bottomrule
\end{tabularx}}
\hfill{}
\caption{Comparison of properties of systems for geospatial link discovery}
\label{tab:geospatial-linking-methods}
\end{minipage}
\end{table*}
\end{center}

As can be observed in Table~\ref{tab:geospatial-linking-methods},
the LIMES system, that implements \textsc{Radon}, was the one who completed all the link discovery task for all topological relations and performed best for most of them. We, therefore, take LIMES as our main reference point. Nevertheless, LIMES as it is\footnote{We used version 1.5.5, the latest version available at the time of writing}, is not sufficiently flexible to accommodate geospatial data of different formats, and requires external pre-processing of input. Additionally, LIMES assumes an error-free download and curated data-sets, which is not always the case in reality. This motivates us to incorporate advantages of existing techniques in a single solution and test what existing technologies might be used for an efficient, flexible, robust and interoperable system for on-the-fly semantic linking of geospatial data.

\section{Geo-L}
We developed a system for geo-spatial linking, which provides the required functionality and shows high performance and accuracy. Geo-L also offers flexible configuration options for the SPARQL affine user as well as accurate error handling.   

\subsection{Input}
The input for a link discovery task provides the resources to be linked and the conditions upon which the links are generated, in a simple, yet flexible manner. In particular, our method offers a way to retrieve relevant properties from the endpoint via a SPARQL query; thus it natively supports manipulation of data, without any need for external pre-processing.
This is useful, for example, when geometry values at the endpoint are not represented in a format that directly allows computations of topological relations.

\subsection{Download}
Downloading from a SPARQL endpoint might occasionally be interrupted before the complete dataset has been delivered. To avoid a total loss of the data our solution does not store all the data in memory while downloading, but instead, periodically write smaller chunks to disk.  
In addition, download might take a relatively long time due to application implementation itself. Our solution seeks to improve this state by reducing the application overhead when querying the remote endpoint.

\subsection{Caching}
To accelerate access to the source- and target-resources we incorporate a caching mechanism. Data retrieved from the SPARQL endpoint are stored in a central data store with an internal index. Further requests for data items from the same endpoint will be first served from the cache if the items are already indexed. This ensures a single local resource parallel to the endpoint, which serves arbitrarily many configurations, thus saves both time and storage.  This differs from the behavior of LIMES, where data items may be downloaded multiple times, and duplicates of the data may be then stored. 
\Cref{alg:caching} sketches the caching process. The method essentially compares the required triples range to the triple indices stored in an internal database,  based on offset and limit parameters given in the configuration. It detects the indices of triples which are not already stored, retrieves the respective triples in chunks from the endpoint, and stores them in the database.    

\begin{algorithm*}
\DontPrintSemicolon
  \SetKwProg{Fn}{Function}{:}{}
  \KwInput{resource.endpoint, resource.id resource.geo, config.offset, config.limit  }
  
  \tcc{create table if not exists for resource}
  T $\leftarrow$ get-table(resource,DB) \; 
  \If{\begin{math} T \not\in DB\end{math}}
  {
  T $\leftarrow$  create-table(resource.id,resource.geo, server-offset) \;
  create GIST-index(T.geo)\;
  }
  \tcc{checking cache}
  min-offset $\leftarrow$ offset\;
  [min-server-offset,max-server-offset] $\leftarrow$ get-stored-offsets()\; 
        max-offset $\leftarrow$ offset + limit - 1\;
        \uIf(\tcp*[h]{all queried records are before all stored ones }){min-server-offset > 0 \AND max-offset < min-server-offset} 
        {
            \tcc{download triples with from the given offset, save after every chunk }
            retrieve-triples(resource.endpoint, offset,limit,chunksize,T)\; 
            
        }
        \uElseIf(\tcp*[l]{all queried records are after all stored ones }){ min-offset > max-server-offset}
        {
            \If(\tcp*[h]{are there any more entries at this offset?  }){endpoint-has-more-entries(min-offset + 1)}
            {
                retrieve-triples(resource.endpoint, offset,limit,chunksize,T)\;
            }          
        }
        \Else(\tcp*[h]{queried entries and stored entries overlap })
        {
            \tcc{find intervals of triple indices to be downloaded}
            intervals = list() \tcp*[h]{list of range pairs i.e., (start-range, end-range)  }\;
            \If{offset < min-server-offset}
            {
                interval.append((offset, min-server-offset - 1))\;
                min-offset = min-server-offset\;
                
            }
            
            \If{max-offset > max-server-offset}
            {
                \If{endpoint-has-more-entries(max-server-offset + 1)}
                {
                    intervals.append((max-server-offset + 1,max-offset))\;
                }
                max-offset $\leftarrow$ max-server-offset\;
                
            }
            
            missing-limit = max-offset - min-offset + 1\;
            \tcc{find intervals of triple indices to be downloaded}
            missing-intervals = find-missing-data(min-offset, missing-limit)\;
            missing-intervals = \CONCAT(intervals, missing-intervals)\;
            \If{length(missing-intervals) > 0}
            {
                \ForEach{interval $\in$ missing-intervals}
                {
                    interval-offset $\leftarrow$ interval[0]\;
                    interval-limit  $\leftarrow$ interval[1] - interval-offset + 1\;
                    retrieve-triples(resource.endpoint, interval-offset,interval-limit,chunksize,T)
                
                }
            }
            
        }
        

\caption{Dataset Caching}\label{alg:caching}
\end{algorithm*}

%
\subsection{Link Discovery}
The task of geo-links discovery requires efficient processing of spatial data, and therefore we use R-trees \citep{guttman1984rtrees} as our underlying data structure.
An R-tree is a data structure used to store and query multi-dimensional objects, in a way that and preserves spatial relations, as vicinity and nesting, among the indexed objects. An R-tree represents each object by its minimum bounding box (MBB), i.e., the smallest rectangle that encloses it, and a leaf node stores the MBB of that object and a pointer to the actual geometry.
An R-tree is organized hierarchically; it groups MBBs by proximity and represents them by \emph{their} MBB in a higher level of the tree.
This process proceeds until all the MBBs are nested in a single bounding box - the tree root.   R-Trees have shown to be efficient in processing spatial joins, to find topological relations between different data sets \citep{brinkhoff1993spatial-joins-r-trees}.
R-Trees support both individual elements search as well as range search, where all the items within a rectangle are retrieved.

A practical problem occurs when the data contain errors, i.e., invalid geometries. The implications of using such data are wrong results, application performance issues, etc. For this reason, geometries are examined before indexing; invalid geometries are not indexed, and thus do not participate in the link discovery.

\subsection{Implementation}
We use Python as our preferred programming language, since it became the language of choice for data science in general, and provides useful tools for handling geospatial data, in particular. We have experimented with the following technologies:

\subsubsection{GeoPandas}
Our initial implementation involved custom built caching and mapping mechanisms. 
We use Python's GeoPandas library \citep{joradahl2017geopandas}, which implements data structures for storing geometric types, as well as analysis tools for geospatial data.
In particular, GeoPandas provides an interface for spatial joins, which allow combining observations stored in these data structures based on their spatial relations. For this purpose GeoPandas indexes geometries using R\textsuperscript{*}-Tree \citep{beckmann1990r-star-tree},  a variant of R-Tree that provides better search performance, at the cost of increased construction time.  GeoPandas  currently supports finding the following spatial relations: \emph{within}, \emph{intersects}, and \emph{contains}.   

We further experimented with cython \citep{behnel2011cython}, a language which is a superset of Python, where code can be compiled directly to C, generating efficient code. 
GeoPandas has been reimplemented in Cython in a way that optimizes the storage of geometries and should improve the performance of spatial operations.

\subsubsection{PostgreSQL}
Furthermore, we implement the system using PostgreSQL, an open source object-relational DBMS, with PostGIS extension, which provides functionality to manage geospatial data, such as geometry data types, efficient indexing, and spatial joins, and is compliant with the Open Geometry Consortium  (OGC) OpenGIS specifications. PostGIS implements spatial indexing with an R-Tree-over-GiST \citep{postgis2.5.0}. GiST, Generalized Search Tree \citep{hellerstein1995generalized}, is a height-balanced tree structure and allows arbitrary indexing schemes.
The choice to use this as the backend of our is multi-fold:
\begin{itemize}
    \item GiST indexes are ``null safe'', therefore attempting to build an R-Tree on data which contains an empty geometry field will fail.
    \item GiST uses a compression technique which results in fast indexing.
    \item The database facilitates the implementation of the resource caching mechanism
\end{itemize}
The source code of Geo-L is available at \url{https://github.com/DServSys/Geo-L}

\section{Experimental Settings}

\subsection{Datasets}
\label{sec:datasets}
The evaluation has been done by finding different relations between points to polygons, and polygons to polygons in the following datasets.

\begin{itemize}
    \item SPOI - Smart Points of Interest: A data set, which contains over 30 million Points of Interest important for tourism around the world \citep{cerba2016spoi}. 
    \item OLU - Open Land-Use: Maps land use on local and regional level; contains over 11 million  geometries -- Polygons and MultiPolygons \citep{mildorf2014open}.
    \item NUTS - Nomenclature of Territorial Units for Statistics: A standard for referencing European countries and their regions, for statistical processes \citep{eurostat2015regions}.
\end{itemize}
These datasets are stored under different graphs in the SPARQL enpoint\footnote{\url{https://www.foodie-cloud.org/sparql}} of the FOODIE project\footnote{\url{http://www.foodie-project.eu/}}.  
While SPOI and OLU are excellent examples for big (open) linked data, NUTS is a standard schema. 
NUTS geometries are not represented in WKT form, and must be be manipulated to conform to the form required by procedures of topological relations computation.
Tools like LIMES, however, do no support such cases.

We compare the performance of LIMES and Geo-L with respect to both topological relations discovery and data retrieval time from endpoints.

\subsection{Experiments}
The performance of the Geo-L systems is evaluated in terms of runtime by conducting experiments on simulations test-sets as well as real-world scenarios. We also note differences in linking results if they occur.  In order to compare the performance of our system with that LIMES, which is implemented using parallel processing.
The task is viewed as consisting of two stages: download and caching, and linking; we report the performance for each of them.  
The simulations enable evaluation of system performance under realistic conditions, with scenarios which otherwise might not be explored, and at the same time providing reliable way confirm their results.
All experiments have been performed on a 64-bit Linux machine with an Intel Core i7-7800X CPU @ 3.50GHz and total of 12 threads (6 CPU cores $\times$ 2 threads per core).   

\subsubsection{Simulation}
Our simulations consist of finding topological relations where the subsets of OLU dataset are used as both source and target datasets. This setting has multiple advantages: First, it allows to demonstrate the benefits of caching, regarding data sets retrieval. 
Additionally, the structure of the OLU set, which consists of separate geometries with non-hierarchical relations, facilitates the link quality evaluation.
We used this approach to perform a preliminary comparison of three implementations on a subset of 165,000 entities (as source and target sets) and observed that the implementations which used GeoPandas performed considerably slower than the one which employed PostgreSQL with PostGIS. For example, the mapping time required for calculating the \emph{within} relation was 38 seconds for the implementation which used GeoPandas, about 20 minutes for the GeoPandas cython implementation, and less than 4 seconds for the implementation which used PostgreSQL.    Therefore, in the following experiments, the latter serves as our reference system.

\begin{figure}
\begin{minipage}{\columnwidth}
\centering
\begin{tikzpicture}
  \begin{axis}[
        ybar, axis on top,
        height=7.5cm,  width  = 0.75\textwidth, 
        bar width=0.2cm,
        ymajorgrids=true, tick align=inside,
        major grid style={draw=white},
        enlarge y limits={value=.1,upper},
        ymin=0, 
        axis x line*=bottom,
        axis y line*=left,
        y axis line style={opacity=0},
        tickwidth=0pt,
        enlarge x limits=true,
        axis y discontinuity=crunch,
        legend style={
            at={(0.5,-0.2)},
            anchor=north,
            legend columns=-1,
            /tikz/every axis label/.append style={font=\footnotesize}, 
            /tikz/every  column /.append style={column sep=0.7cm},
        },
        xlabel={Instances},
        ylabel={Retrieval time (s)},
        symbolic x coords={
           $165\cdot10^3\times165\cdot10^3$,
           $400\cdot10^3\times400\cdot10^3$,
           },
       xtick=data,
       nodes near coords,
       every node near coord/.append style={rotate=90,anchor=west, font=\footnotesize},
    ]
    \addplot [draw=none,fill=red!50] coordinates {
      ($165\cdot10^3\times165\cdot10^3$, 37.053)
      ($400\cdot10^3\times400\cdot10^3$, 88.545)
       };
   \addplot [draw=none, fill=green!60!black] coordinates {
      ($165\cdot10^3\times165\cdot10^3$, 18.690)
      ($400\cdot10^3\times400\cdot10^3$, 13.325) 
       };

    \legend{LIMES,Geo-L}
  \end{axis}
  \end{tikzpicture}
\caption{Retrieval time OLU-OLU}
\label{fig:olu_olu_retrieval}
\end{minipage}
\end{figure}
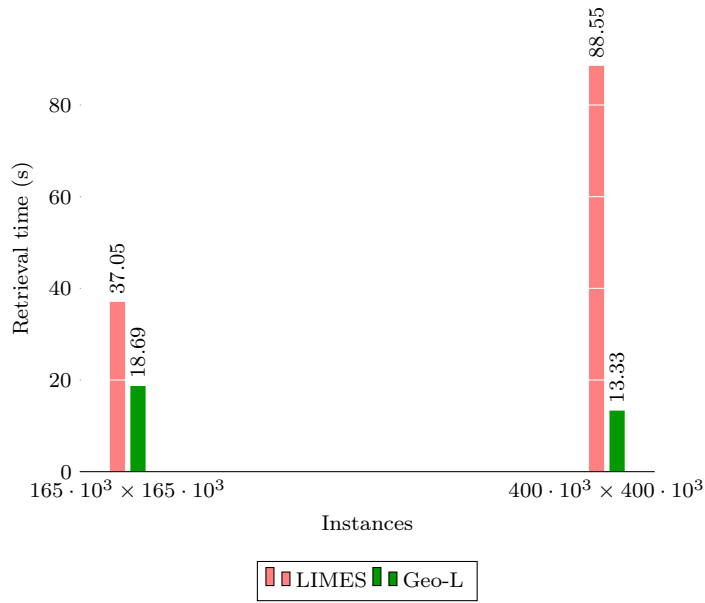

We tested the systems with two subsets: the one contains the first 165,000 geometries, and the other the first 400,000 geometries.
Figure~\ref{fig:olu_olu_retrieval} compares the dataset the retrieval times of OLU subsets for both LIMES and Geo-L. The first scenario shows that retrieval time for LIMES is about twice as long compared to Geo-L. The reason is that LIMES does not detect whether data already exist, and download the same OLU subset twice, both as source and target datasets.
The second scenario emphasizes this phenomenon: Whereas Geo-L retrieves the data which has not already been downloaded yet, and does it only once, LIMES retrieves twice the subset of 400,000 geometries, which takes more than six times longer.


Moreover, LIMES stores redundant data e.g., the subset of the first 165,000 geometries is store four times, as it is contained in the  400,000 geometries subset.  

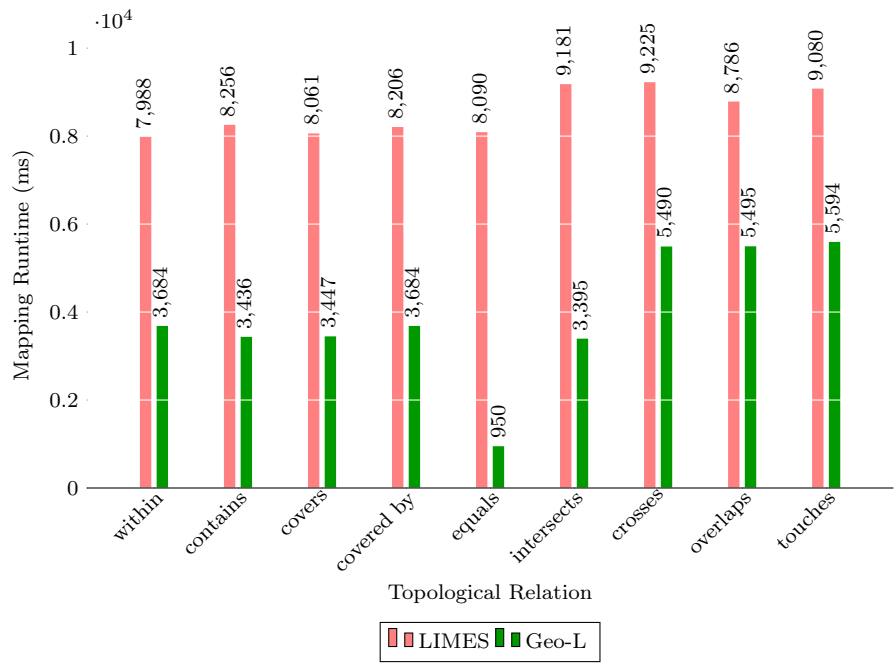
\begin{figure}
\begin{minipage}[b]{\columnwidth}
\begin{tikzpicture}
  \centering
  \begin{axis}[
        ybar, axis on top,
        height=7.5cm,  width  = \textwidth, 
        bar width=0.15cm,
        ymajorgrids=true, tick align=inside,
        major grid style={draw=white},
        enlarge y limits={value=.1,upper},
        ymin=0, 
        axis x line*=bottom,
        axis y line*=left,
        y axis line style={opacity=0},
        tickwidth=0pt,
        enlarge x limits=true,
        axis y discontinuity=crunch,
        legend style={
            at={(0.5,-0.3)},
            anchor=north,
            legend columns=-1,
            /tikz/every axis label/.append style={font=\footnotesize},
            /tikz/every  column /.append style={column sep=3cm},
        },
        xlabel={Topological Relation},
        label style={font=\footnotesize},
        x tick label style={rotate=45, anchor=north east, inner sep=0mm,font=\footnotesize},
        ylabel={Mapping Runtime (ms)},
        symbolic x coords={
           within,contains,covers,covered by
           ,equals, intersects, crosses, overlaps, touches
           },
       label style={font=\footnotesize},       
       xtick=data,
       nodes near coords,
       every node near coord/.append style={rotate=90,anchor=west,font=\footnotesize},
    ]
    \addplot [draw=none,fill=red!50] coordinates {
      (within, 7988)
      (contains, 8256) 
      (covers, 8061)
      (covered by,8206) 
      (equals,8090)
      (intersects,9181)
      (crosses,9225) 
      (overlaps,8786)
      (touches,9080)
       };
   \addplot [draw=none, fill=green!60!black] coordinates {
      (within, 3684)
      (contains, 3436) 
      (covers, 3447)
      (covered by,3684) 
      (equals,950)
      (intersects,3395)
      (crosses,5490)
      (overlaps,5495)
      (touches,5594)
       };

    \legend{LIMES,Geo-L}
  \end{axis}
  \end{tikzpicture}
\caption{Performance OLU-OLU; size: $165\cdot10^3\times165\cdot10^3$ }
\label{fig:olu165K_olu165K}
\end{minipage}
\end{figure}

\begin{figure}
\begin{minipage}[b]{\columnwidth}
\begin{tikzpicture}
  \centering
  \begin{axis}[
        ybar, axis on top,
        height=7.5cm,  width  = \textwidth, 
        bar width=0.15cm,
        ymajorgrids=true, tick align=inside,
        major grid style={draw=white},
        enlarge y limits={value=.1,upper},
        ymin=0, 
        axis x line*=bottom,
        axis y line*=left,
        y axis line style={opacity=0},
        tickwidth=0pt,
        enlarge x limits=true,
        axis y discontinuity=crunch,
        legend style={
            at={(0.5,-0.3)},
            anchor=north,
            legend columns=-1,
            /tikz/every axis label/.append style={font=\footnotesize},
            /tikz/every  column /.append style={column sep=3cm},
        },
        xlabel={Topological Relation},
        label style={font=\footnotesize},
        x tick label style={rotate=45, anchor=north east, inner sep=0mm,font=\footnotesize},
        ylabel={Mapping Runtime (ms)},
       symbolic x coords={
           within,contains,covers,covered by, 
           equals, intersects, crosses, overlaps, touches
          },
       xtick=data,
       nodes near coords,
       every node near coord/.append style={rotate=90,anchor=west,font=\footnotesize},
    ]
    \addplot [draw=none,fill=red!50] coordinates {
      (within, 17227)
      (contains, 17390) 
      (covers, 17189)
      (covered by,17460) 
      (equals,15608)
      (intersects,22955)
      (crosses,22295) 
      (overlaps,22430)
      (touches,22311)
       };
   \addplot [draw=none, fill=green!60!black] coordinates {
      (within, 10913)
      (contains, 8837) 
      (covers, 8846)
      (covered by,10903) 
      (equals,2143)
      (intersects,9114)
      (crosses,20684)
      (overlaps,20696)
      (touches,21190)
       };
    \legend{LIMES,Geo-L}
  \end{axis}
  \end{tikzpicture}
\caption{\small Performance OLU-OLU; size: $400\cdot10^3\times400\cdot10^3$ }
\label{fig:olu400K_olu400K}
\end{minipage}
\end{figure}
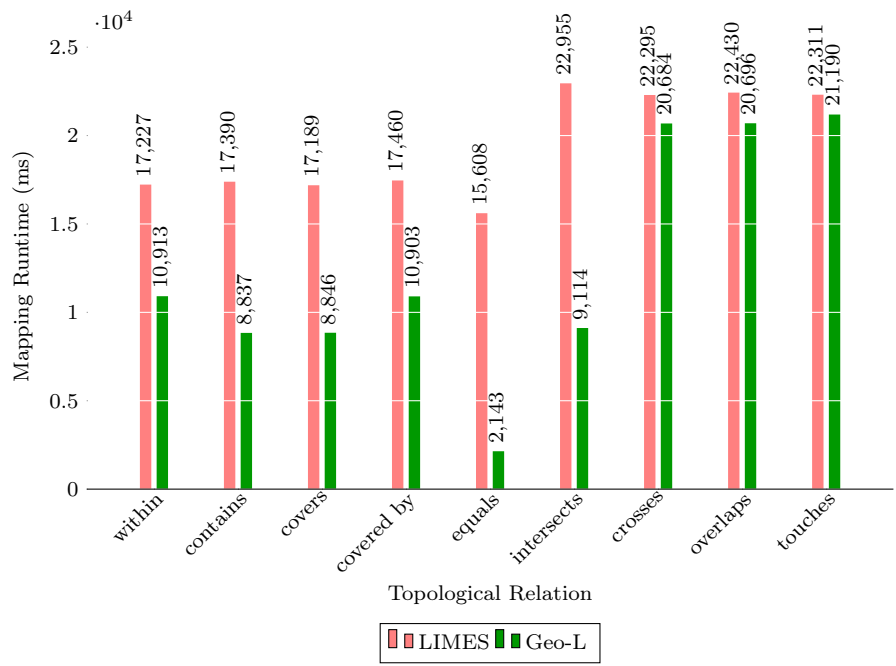

Experiments have been repeated ten times for each topological relation type per subset, and the average mapping times are shown for both LIMES and Geo-L in Figure~\ref{fig:olu165K_olu165K} and Figure~\ref{fig:olu400K_olu400K}.
As can be observed, Geo-L discover topological links faster than LIMES, for all relations in these experiments.
The coefficients of variation (CV) of runtimes for the different experiments 
were found to be low in all cases (CV $<$ 0.1), which indicates that these results are consistent.   

In addition, we found discrepancies between the links discovered by each system. 
For example, when looking for links of entities which stand in the \texttt{within} relation in two sets with identical entities, the expected result is that each item in the source set would stand in this relation with exactly one entity of the target set, and that the size of the returned set would be equal to the size of each of sets. However, for the $165 \cdot  10^3$ OLU subset Geo-L found  $164,935$ links, whereas LIMES found $155,083$.
The  65 entities which Geo-L did not include had invalid geometries, which were detected already during construction and omitted from the search space.  
We examined the result computed by LIMES and noticed that the difference of $9852$ consisted mostly of ``false negatives'' errors, i.e., valid geometries which were omitted from the result set ($9849$ links). Also, there were three links that Geo-L did not found and LIMES did. These, however, are ``false positives'', i.e., the links contained invalid geometries, which were included in the result set by LIMES, whereas Geo-L has omitted them  already before computing the links.
Similar errors occurred also for other topological relations.

\subsubsection{Real-World Scenarios}
We experiment with topological relation discovery between pairs of geospatial resources mentioned in Section~\ref{sec:datasets}, and compare their performance to that of LIMES. Figure~\ref{fig:spoi_olu} shows the performance, in terms of mapping runtime, on different subsets of SPOI and OLU. In this example the largest subset does not contain the other two: the first $500\cdot10^3$ entities of OLU contain geometries which caused  LIMES system to crash, and therefore we chose a subset of the same size but specified a different offset.

Figure~\ref{fig:spoi_nuts} shows the running times for mapping SPOI to NUTS with different subset sizes of SPOI. Since NUTS geometries are not represented in WKT format we used a configuration feature which defines a resource via a SPARQL query. In this case, the query also transforms the geometries into the required format. This, however, is not possible in LIMES, and therefore comparison of the systems is not presented.

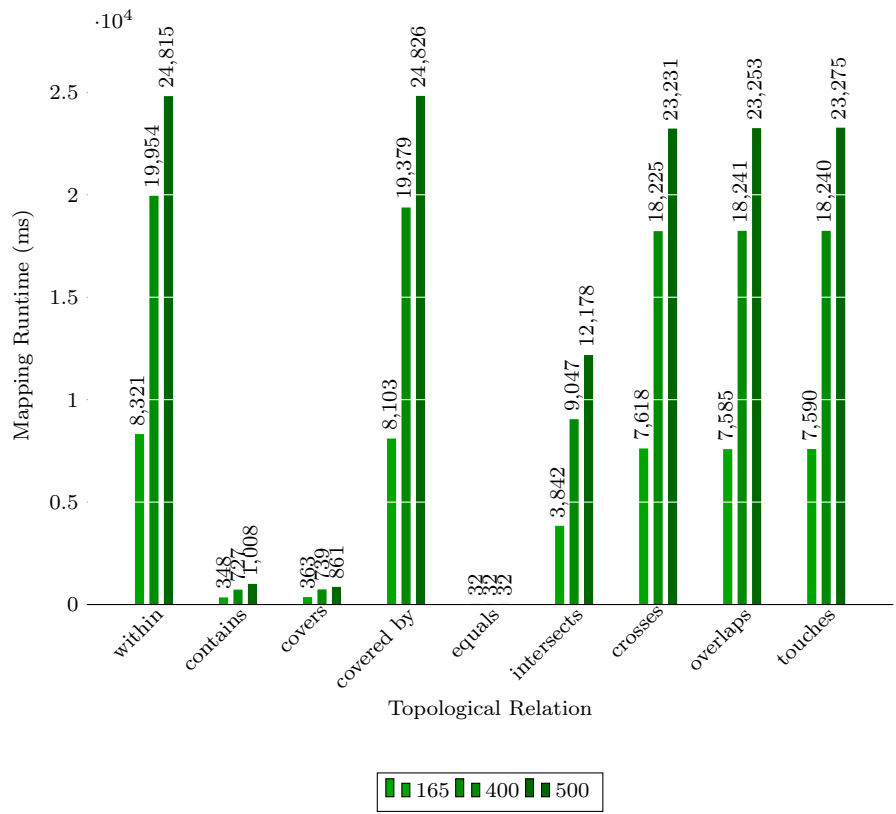
\begin{figure}[ht]
\begin{minipage}[b]{\columnwidth}
\begin{tikzpicture}
  \centering
  \begin{axis}[
        ybar, axis on top,
        height=9cm,  width  = \textwidth, 
        bar width=0.12cm,
        ymajorgrids=true, tick align=inside,
        major grid style={draw=white},
        enlarge y limits={value=.1,upper},
        ymin=0, 
        axis x line*=bottom,
        axis y line*=left,
        y axis line style={opacity=0},
        tickwidth=0pt,
        enlarge x limits=true,
        axis y discontinuity=crunch,
        legend style={
            at={(0.5,-0.3)},
            anchor=north,
            legend columns=-1,
            /tikz/every axis label/.append style={font=\footnotesize},
            /tikz/every  column /.append style={column sep=2.5cm},
        },
        xlabel={Topological Relation},
        x tick label style={rotate=45, anchor=north east, inner sep=0mm, font=\footnotesize},
        ylabel={Mapping Runtime (ms)},
        label style={font=\footnotesize},
        symbolic x coords={
           within,contains,covers,covered by, 
           equals, intersects, crosses, overlaps, touches
           },
       xtick=data,
       nodes near coords,
       every node near coord/.append style={rotate=90,anchor=west},
    ]
  \addplot [draw=none, fill=green!65!black] coordinates {
      (within, 8321)
      (contains, 348) 
      (covers, 363)
      (covered by,8103) 
      (equals,32)
      (intersects,3842)
      (crosses,7618)
      (overlaps,7585)
      (touches,7590)
       };
    \addplot [draw=none, fill=green!55!black] coordinates {
      (within, 19954)
      (contains, 727) 
      (covers, 739)
      (covered by,19379) 
      (equals,32)
      (intersects,9047)
      (crosses,18225)
      (overlaps,18241)
      (touches,18240)
       };
    \addplot [draw=none, fill=green!40!black] coordinates {
      (within, 24815)
      (contains, 1008) 
      (covers, 861)
      (covered by,24826) 
      (equals,32)
      (intersects,12178)
      (crosses,23231)
      (overlaps,23253)
      (touches,23275)
       };   
    \legend{165,400,500}
  \end{axis}
  \end{tikzpicture}
\caption{Performance OLU-NUTS; size: $X\cdot10^3\times1782$ }
\label{fig:olu_nuts}
\end{minipage}
\end{figure}

Figrue~\ref{fig:olu_nuts} shows mapping runtime for different subsets of OLU to NUTS, for different topological relations.

\begin{figure}[ht]
\begin{minipage}[b]{\columnwidth}
\begin{tikzpicture}
  \centering
  \begin{axis}[
        ybar, axis on top,
        height=6.5cm,  width  = \textwidth, 
        bar width=0.2cm,
        ymajorgrids=true, tick align=inside,
        major grid style={draw=white},
        enlarge y limits={value=.1,upper},
        ymin=0, 
        axis x line*=bottom,
        axis y line*=left,
        y axis line style={opacity=0},
        tickwidth=0pt,
        enlarge x limits=true,
        axis y discontinuity=crunch,
        legend style={
            at={(0.5,-0.2)},
            anchor=north,
            legend columns=-1,
            /tikz/every axis label/.append style={font=\footnotesize},
            /tikz/every  column /.append style={column sep=0.7cm},
        },
        xlabel={Instances},
        ylabel={Mapping Runtime (ms)},
        symbolic x coords={
           $165\cdot10^3\times165\cdot10^3$,
           $400\cdot10^3\times400\cdot10^3$,
           $500\cdot10^3\times500\cdot10^3$
           },
       xtick=data,
       label style={font=\footnotesize},
       nodes near coords,
       every node near coord/.append  style={rotate=90,anchor=west, font=\footnotesize},
    ]
    \addplot [draw=none,fill=red!50] coordinates {
      ($165\cdot10^3\times165\cdot10^3$, 2862)
      ($400\cdot10^3\times400\cdot10^3$, 5978)
      ($500\cdot10^3\times500\cdot10^3$, 13885)
       };
   \addplot [draw=none, fill=green!60!black] coordinates {
  
      ($165\cdot10^3\times165\cdot10^3$, 810)
      ($400\cdot10^3\times400\cdot10^3$, 1839) 
      ($500\cdot10^3\times500\cdot10^3$, 2171)
       };

     \legend{LIMES,Geo-L}
  \end{axis}
  \end{tikzpicture}
\caption{Performance SPOI-OLU; topological relation:  within}
\label{fig:spoi_olu}
\end{minipage}
\begin{minipage}[b]{\columnwidth}
\begin{tikzpicture}
  \centering
  \begin{axis}[
        ybar, axis on top,
        height=6.5cm,  width  = \textwidth, 
        bar width=0.2cm,
        ymajorgrids=true,  tick align=inside,
        major grid style={draw=white},
        enlarge y limits={value=.1,upper},
        ymin=0, 
        axis x line*=bottom,
        axis y line*=left,
        y axis line style={opacity=0},
        tickwidth=0pt,
        enlarge x limits=true,
        axis y discontinuity=crunch,
        legend style={
            at={(0.5,-0.2)},
            anchor=north,
            legend columns=-1,
            /tikz/every axis label/.append style={font=\footnotesize},
            /tikz/every  column /.append style={column sep=0.7cm},
        },
        xlabel={Instances},
        ylabel={Mapping Runtime (ms)},
        symbolic x coords={
           $165\cdot10^3\times1782$,
           $400\cdot10^3\times1782$,
           $500\cdot10^3\times1782$
           },
       label style={font=\footnotesize},       
       xtick=data,
       nodes near coords,
       every node near coord/.append style={rotate=90,anchor=west, font=\footnotesize},
    ]
   \addplot [draw=none, fill=green!60!black] coordinates {
      ($165\cdot10^3\times1782$, 753)
      ($400\cdot10^3\times1782$, 1432)
      ($500\cdot10^3\times1782$, 2044)
       };

     \legend{Geo-L}
  \end{axis}
  \end{tikzpicture}
\caption{Performance SPOI-NUTS; topological relation:  within}
\label{fig:spoi_nuts}
\end{minipage}
\end{figure}

The system has been employed as part of DataBio, a EU Horizon 2020 project. A main goal of the project is to show the benefits of Big Data technologies in the raw material production from agriculture for the bioeconomy industry. The project uses Linked Data as a federated layer to integrate to integrate cross-organizational heterogeneous data.

In particular, Geo-L has been successfully applied to various use cases in field management, e.g.:
\begin{itemize}
    \item identifying plots from the Czech registry of farmland, which intersect with \emph{buffer zones} around water bodies. 
    A buffer zone is a vegetated or forested strip around lakes and along water courses. Its purpose, in the context of agricultural management, is to protect water bodies from pollutants as pesticides, nutrients, and sediment \citep{zhang2010review}. Therefore, it is crucial to detect cases where field areas and buffer zones intersect. 
    \Cref{fig:buffer-zone} depicts a case where a buffer zone of a lake intersects with a field, marked with orange 
    \begin{figure}[ht]
      \includegraphics[scale=0.38173]{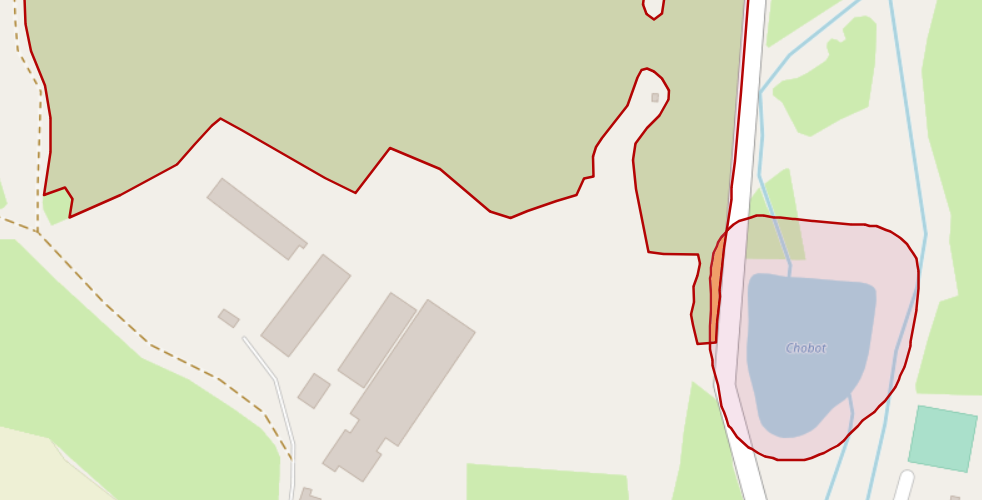}
      \caption{Buffer zone of a lake which intersects with a field }
      \label{fig:buffer-zone}
    \end{figure}
    \FloatBarrier
    \item identifying erosion zones for a specific farm. Soil erosion zones is the detachment and deposition of soil particles. It may be caused by e.g., wind, snow, water, but also due to human-induced land use   \citep{vanwalleghem2016soil}. As the latter results in much faster erosion rates it can effect soil quality dramatically due to loss of nutrients as well as the ability to get them.  It is therefore important to control erosion This, since it impacts productivity and sustainability negatively \citep{larson1983threat,blanco2010erosion}. \Cref{fig:erosion-zones} shows erosion zones overlap with a plot, marked in dark blue.
    \begin{figure}[ht]
      \includegraphics[scale=0.45]{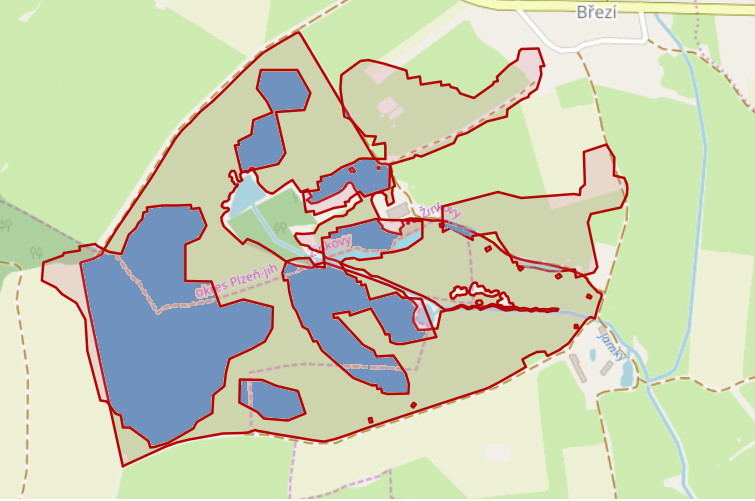}
      \caption{Erosion zones land of a field }
      \label{fig:erosion-zones}
    \end{figure}
    \FloatBarrier
    \item identifying fields within a particular region, which grow the same crop type for a specific year as given field in that region. This serves as an assisting tool for farm management and agricultural landscape planning, e.g.,  controlling crop  diversification or rotation. \Cref{fig:sillage-maize} presents all fields, which grow the same crop type like the field marked in brown, here, maize for silage, during 2019, within the South Moravian Region (region border marked in grey).
    \begin{figure}[ht]
      \includegraphics[scale=0.4]{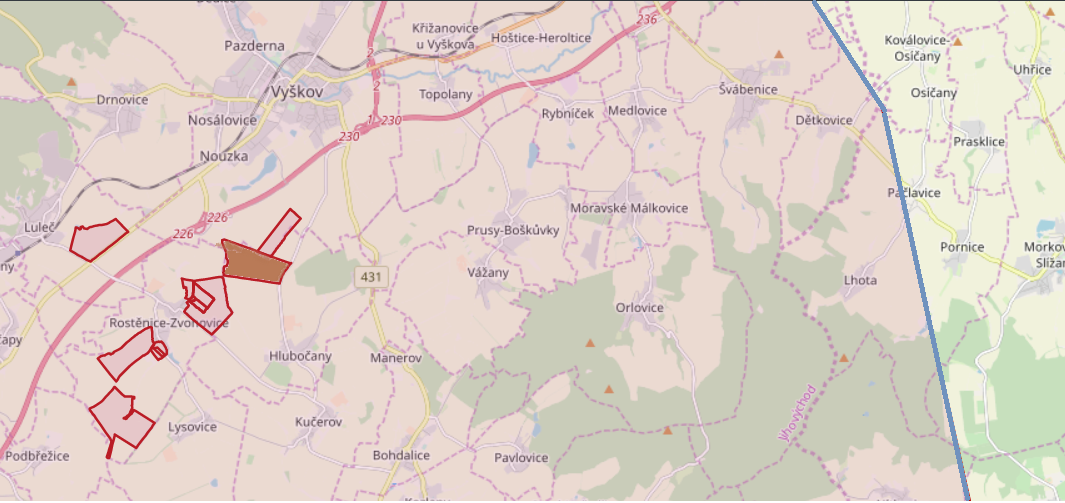}
      \caption{Fields which grow maize for silage during 2019 within the South Moravian Region}
      \label{fig:sillage-maize}
    \end{figure}
    
\end{itemize}

\section{Conclusions}
This paper presented Geo-L, a system for discovering RDF links between geospatial entities, based on topological relations.
We conducted experiments to detect topological relations between points and polygons, and between polygons and polygons.
The experiments show that Geo-L outperforms LIMES \citep{ngomo2011limes}, a state-of-the-art link discovery system, for this task in several aspects:  
\begin{itemize}
    \item Scalability and efficiency: Geo-L configuration allows to form a dataset directly by the SPARQL query that defines it. This feature is, in particular, useful when data at the SPARQL endpoint are stored differently than specified for the linking task, but could be transformed  into the required format through SPARQL functions.
    \begin{itemize}
        \item Download time: Datasets are cached not for a single task but are regarded as resources of their own. Thanks to its caching mechanism, Geo-L accesses the SPARQL endpoints only when data required in the dataset are missing, and expands existing datasets where possible.
        \item Mapping time: Geo-L utilizes PostgreSQL with PostGIS index for storing and indexing of the data. This enables efficient spatial joins between source- and target-datasets.
    \end{itemize}
    \item Robustness: Geo-L includes multiple features that strengthen the robustness of the application.  
    \begin{itemize}
        \item Caching: Geo-L caches portions of the data as they are downloaded, rather than writing the whole dataset after being downloaded. This property prevents data loss when, e.g., connection to the remote endpoint is lost.  
        \item Mapping accuracy: Geo-L detects entities with invalid geometries (compliant to OGC OpenGIS specification) and does not include them in the search space. In addition, in several cases LIMES did not include valid geometries in the result set, whereas Geo-L correctly did.
    \end{itemize}
    \item Interoperability and flexibility: Geo-L can be used as a stand-alone application or as a REST service (in a docker), which allows it to be integrated with other applications. The easy SPARQL-based and slim set-up of target and source configuration (as JSON) enables a very free usage of the tool. 
\end{itemize}

Future work will examine relations between other types of geometries as well as explore geospatial relations based on various distance measures.
The current implementation recalls the same items for each dataset once they are cached. In the future we will also address re-caching in case data at the SPARQL endpoint have been modified, an issue which is, to the best of our knowledge, not handled by other geospatial-linking systems.

\bibliographystyle{spbasic}      

\bibliography{bibliography} 

\end{document}